# Kinetic Processes to Radio Burst: First Observational-driven Study in Coronal Loops

Mehdi Yousefzadeh,[1] Marian Karlický,[2] Artem Koval,[2] Alena Zemanova,[2] Yao Chen,[3,4] and Jun Lin[1,5,6]

[1] *Yunnan Observatories, Chinese Academy of Science, Kunming, Yunnan 650216, People's Republic of China*
[2] *Astronomical Institute of the Czech Academy of Sciences, Fričova 298, CZ-25165 Ondřejov, Czech Republic*
[3] *Institute of Frontier and Interdisciplinary Science, Shandong University, Qingdao, Shandong 266237, People's Republic of China*
[4] *School of Space Science and Physics, Shandong University, Weihai, Shandong 264209, People's Republic of China*
[5] *University of Chinese Academy of Sciences, Beijing 100049, People's Republic of China*
[6] *Yunnan Key Laboratory of Solar Physics and Space Science, Kunming 650216, People's Republic of China*

## ABSTRACT

Understanding the origin of coherent solar radio bursts requires linking macroscopic coronal structures with the kinetic processes responsible for wave generation. We investigate emissions from slowly positively drifting bursts (SPDBs), a specific type of solar radio emission. SPDBs provide observational constraints for modeling beam-plasma interactions in coronal loops, serving as a basis for a multiscale description. We employ a three–stage numerical framework that combines nonlinear force–free field (NLFFF) magnetic extrapolation, guiding–center simulations, and fully kinetic particle–in–cell (PIC) modeling. The background plasma density is described using a hydrostatic model consistent with active–region conditions, producing a plasma–frequency gradient comparable to that inferred from the observed SPDBs spectrum. Energetic electrons injected near the loop top evolve through magnetic mirroring, pitch–angle scattering, turbulence development, and partial precipitation. Evolved velocity distribution functions (EVDFs) are sampled after approximately one bounce period and used in PIC simulations to evaluate emission properties. The results show that the evolved beam distribution of energetic electrons predominantly excites beam–Langmuir waves and fundamental plasma emission along the loop, with the emission intensity gradually decreasing from the loop top toward the footpoint. The modest initial beam velocity of energetic electrons explains the inefficient generation of harmonic plasma emission. The temporal evolution of the modeled emission reproduces key SPDB characteristics, including the $\sim 4\,\mathrm{s}$ duration and frequency drift behavior. These results suggest plasma emission explains the mechanisms behind SPDB generation and demonstrate the feasibility of a unified model connecting coronal magnetic topology, particle transport, and radio emission.

## 1. INTRODUCTION

Solar flares often produce different types of radio bursts that are key to understanding energetic particle dynamics and plasma processes in the solar corona (Krueger 1979; McLean & Labrum 1985; Aschwanden 2004). Surveys using a wide range of wavelengths have identified different types of radio emissions, and each type is linked to particular processes and areas within the solar atmosphere. In the decimetric frequency domain (0.3–3.0 GHz), a wide variety of bursts have been documented by Guedel & Benz (1988), Isliker & Benz (1994), and Jiřička et al. (2001). Type III and type IV bursts are among the most frequently observed radio bursts in this range. While these two categories dominate solar radio observations, other less frequent but intriguing burst types have also been reported, including

Corresponding author: Mehdi Yousefzadeh
Email: yousefzadeh@ynao.ac.cn

the so-called slowly positively drifting bursts (SPDBs). These SPDBs can possibly offer valuable insights into the processes that can generate radio emissions in the solar atmosphere.

Statistical analyses indicate that SPDBs are frequently associated with solar flares and tend to occur during their early phases (Benz & Simnett 1986; Kotrc et al. 1999; Karlický et al. 2018). These bursts can occur individually or in groups and typically have a linear or slightly curved shape. SPDBs resemble reverse type III bursts, exhibiting a positive frequency drift that indicates propagation into denser layers of the solar atmosphere, likely driven by downward-propagating particle beams. They are characterized by a slow frequency drift toward higher frequencies, reaching up to a few hundred MHz, with durations of one to several seconds, and their origin remains unknown (Jiřička et al. 2001; Karlický et al. 2018; Zemanová et al. 2020). Type III bursts are generally understood as signatures of electron beams propagating through the corona and interplanetary space, producing plasma emission at the local plasma frequency and its harmonic (Aschwanden 2002; Reid & Ratcliffe 2014; Benz 2017). Unlike classical reverse-drift type III bursts, which are attributed to downward-propagating electron beams, SPDBs exhibit significantly smaller drift rates, suggesting different underlying physical processes.

Several mechanisms have been proposed for SPDB generation. Kotrc et al. (1999) suggested downward-propagating shock waves triggered during flare onset as a possible origin, based on radio, H$\alpha$, and spectral imaging data. Karlický (2015)'s numerical simulations indicated that SPDBs could result from radio emission produced by a thermal conduction front propagating through a flaring loop. Supporting this, Karlický et al. (2018) observed a temporal correlation between an SPDB event and the descent of an EUV/UV multithermal plasma blob along a dark H$\alpha$ loop, with corresponding changes in the H$\alpha$ spectral profile. More recently, Zemanová et al. (2020) reported a group of SPDBs that occurred simultaneously with the brightening of a large magnetic rope, suggesting energetic particles propagating along helical magnetic field lines. These observations highlight the complex relationship between flare-associated plasma dynamics, magnetic topology, and radio emission. Despite these advances, the responsible mechanisms and conditions for the generation of SPDB are still not fully understood, and their rarity poses additional challenges for statistical characterization.

The radio bursts are generally attributed to coherent emission mechanisms, wherein plasma instabilities amplify specific wave modes, leading to the generation of intense electromagnetic radiation. One of the main coherent processes responsible for solar and stellar radio bursts is the plasma emission mechanism, which involves the generation and nonlinear evolution of Langmuir waves. These electrostatic oscillations arise when an energetic population of electrons—typically injected during magnetic reconnection events—propagates through a background plasma, driving beam–plasma instabilities once their drift velocity exceeds the local thermal electron speed (e.g., Chen et al. 2022; Zhang et al. 2022; Yousefzadeh 2025).

In this mechanism, Langmuir waves serve as an intermediate step in the conversion of kinetic energy from fast electrons into radio emission. Through nonlinear coupling processes, such as coalescence with ion–acoustic waves or scattering off plasma density fluctuations, Langmuir waves can generate electromagnetic waves at the fundamental ($f \simeq f_{pe}$) and harmonic ($f \simeq 2f_{pe}$) frequencies of the plasma frequency (Ginzburg & Zheleznyakov 1958; Melrose 2017). This mechanism stands in contrast to the Electron Cyclotron Maser Emission (ECME), in which the emission originates from the direct resonance between electrons' cyclotron motion and electromagnetic waves. While both mechanisms are capable of producing highly coherent radiation, the plasma emission process predominates under coronal conditions where $\omega_{pe} \gg \Omega_{ce}$, whereas ECME becomes more efficient when $\Omega_{ce} \gtrsim \omega_{pe}$. Together, these two mechanisms provide complementary insight into the complex interplay between magnetic topology, particle acceleration, and wave–particle interactions in astrophysical plasmas.

To bridge the large-scale and kinetic-scale aspects of radio emission processes, Yousefzadeh et al. (2021) introduced a comprehensive three-step numerical framework. The first step employs the nonlinear force-free field (NLFFF) extrapolation method (Wheatland et al. 2000; Wiegelmann 2004; Wiegelmann et al. 2006) to reconstruct the three-dimensional magnetic configuration of active regions. The second step utilizes the guiding-center approximation (GCA; Northrop 1963) to trace the trajectories and velocity distribution functions (VDFs) of energetic electrons as they propagate along the extrapolated loops. Finally, the third step applies the particle-in-cell (PIC) technique to simulate the kinetic instabilities and wave excitations driven by the nonthermal VDFs derived from the GCA modeling. This multiscale approach establishes a direct link between macroscopic electron transport and microscopic wave–particle dynamics, allowing detailed investigation of the physical processes responsible for the generation of coherent radio emissions.

Using this framework, Yousefzadeh et al. (2021) analyzed the velocity distribution of electrons at the loop-top (LT) region and demonstrated that non-Maxwellian features such as strip-like and loss-cone distributions can efficiently excite the second harmonic (X2) mode via ECME. However, their analysis was confined to the LT region. In a subsequent study, Yousefzadeh et al. (2022) extended this investigation across the entire coronal loop and found that the X2 mode was predominantly excited in the upper loop sections, with the most intense emission near the LT, while the Z mode (i.e., the slow extraordinary mode) showed broader excitation throughout the loop. More recently, Yousefzadeh et al. (2025) examined the effect of the electron injection site—either at the LT or near the loop footpoint (FP)—on wave excitation patterns. Their simulations revealed that LT injection favors electromagnetic (X2 and Z) mode generation, whereas FP injection results in stronger Langmuir wave activity as electrons ascend along the loop.

Building upon these developments, the present study aims to establish a connection between numerical modeling and observational evidence by focusing on a particular class of fine radio structures known as SPDBs. Their highly organized temporal and spectral features suggest that they may originate from periodically modulated particle acceleration or from evolving plasma parameters along magnetic loops. By applying the aforementioned GCA–PIC framework to reproduce conditions consistent with SPDB observations, this work takes the initial steps toward quantitatively linking radio dynamic spectra with kinetic instabilities in realistic coronal environments. Through this combined numerical and observational approach, we aim to advance the understanding of how localized electron injections and loop configurations give rise to the fine spectral structures observed in solar radio emissions.

## 2. RADIO SPECTRUM OBSERVATIONS

In this paper, we refer to the observational study of SPDBs presented by Zemanová et al. (2024). The authors carefully analyzed SPDB radio data obtained by the RT5 radio spectrograph at the Ondřejov Observatory of the Astronomical Institute of the Czech Academy of Sciences. The RT5 operates in the 800–2000 MHz range with a frequency resolution of 4.7 MHz and a temporal resolution of 10 ms (Jiricka et al. 1993; Jiřička & Karlický 2008).

On 2014 May 10, the RT5 recorded three groups of SPDBs during 06:57–07:03 UT. The radio spectrum, with groups SPDB-1, SPDB-2, and SPDB-3, is illustrated in Figure 3a of Zemanová et al. (2024). These bursts occurred during a confined flare of soft X-ray class C8.7. Almost all of the detected radio bursts appeared during the impulsive phase of the flare, except for SPDB-3, which began a few seconds after the flare maximum. These SPDBs resembled reverse type III bursts; however, their frequency drifts of 54–90 MHz $s^{-1}$ were significantly smaller than those typical of type III bursts, $\sim$1 GHz $s^{-1}$, in this frequency range (Aschwanden 2002). Between 06:59:00 and 07:00:20 UT, narrow-band type III bursts were observed in the 900–1400 MHz range.

Figure 1 (upper and lower panels) shows dynamic spectra of SPDBs recorded by the BLEN Callisto instrument (Benz et al. 2005) and RT5 during 07:01–07:02 UT. The BLEN7M Callisto measurements extend the observed frequency range down to 600 MHz and reveal that some bursts originated at frequencies as low as 700 MHz. The frequency drifts measured in SPDBs ranged from 50–180 MHz $s^{-1}$. A summary of all radio bursts detected in the 600–2000 MHz range is provided in Table 1 of Zemanová et al. (2024).

The SPDB event analyzed in this work occurred during the impulsive phase of a solar flare and was previously investigated using a combination of radio, UV/EUV, and hard X-ray observations. Since radio dynamic spectra do not provide direct information about the spatial location of the radio source, the identification of the emitting structure relied on comparing the temporal evolution of the radio flux with light curves extracted from individual flare kernels. These kernels represent compact, bright chromospheric features where sudden upflow movement of plasma and heating take place during solar flares. In particular, the normalized radio flux at 1190 MHz, where the SPDB activity was clearly detected, was compared with UV/EUV emission from several flare kernels distributed throughout the active region. The strongest temporal correlation was found for a kernel located near the spine footpoint of the reconnecting magnetic configuration (denoted K13), suggesting that the SPDB-producing electron beams originated from large-scale magnetic reconnection and subsequently propagated along newly formed flare loops.

Motivated by these observational findings, we initially attempted to model the corresponding large-scale magnetic loop in our NLFFF extrapolation. Although Zemanová et al. (2024) proposed a loop connecting kernels K13 and K3 as a possible source region for the first group of SPDBs, such a magnetic connection could not be reliably recovered in our extrapolated magnetic field. Furthermore, a re-examination of the observational data showed that the temporal evolution of the SDO/AIA 1600 Å and SDO/AIA 304 Å light curves (LCs) at K6 correlates well with the normalized radio flux variations at 1190 MHz associated with the selected SPDB event. Figure 2 displays the approximate locations of K13 and K6 flare kernels over the background HMI image.

We therefore searched for alternative large-scale loops rooted near the observationally constrained source region. Extrapolation identified several candidate loops connecting the vicinity of K13 with the area close to flare kernel K6, where clear magnetic connectivity was present. We selected a loop that satisfies the principal observational constraints of the desired SPDBs that occurred at around 07:01:10 UT, including the approximate source altitude, loop geometry, and inferred electron-beam velocity associated with the observed SPDB frequency drift. Therefore, this loop was adopted as the magnetic structure used as the basis presented in this work.

## 3. NUMERICAL SIMULATION MODELS

The present numerical study extends the three-step modeling framework introduced in previous works (Yousefzadeh et al. 2021, 2022, 2025), with key modifications designed to incorporate observational constraints derived from SPDBs. This approach establishes a consistent link between large-scale coronal magnetic topology, plasma parameters inferred from the observed radio spectrum, and the microscopic kinetic processes responsible for coherent radio emissions.

### 3.1. *Magnetic Field Extrapolation and Plasma Density Model*

The magnetic configuration analyzed in this study corresponds to the active region NOAA 12056, which was associated with the observed SPDB event. The large-scale magnetic field structure was reconstructed using the NLFFF extrapolation method (Wheatland et al. 2000; Wiegelmann 2004; Wiegelmann et al. 2006), applied to photospheric vector magnetograms obtained by the Helioseismic and Magnetic Imager (HMI; Schou et al. 2012). The extrapolated field line of the selected coronal loop derived from the full disk magnetogram data observed at 07:00 UT is shown in Figure 2(c), representing the region where the emission was likely generated.

To derive a plasma density profile consistent with the radio–emission frequency range observed in the SPDB dynamic spectrum, we adopt a hydrostatic density model along the reconstructed magnetic field line derived from the NLFFF extrapolation. The SPDB observation spans the 600-1100 MHz frequency range (Figure 1), corresponding to a density range of $4.44\times10^9\,\mathrm{cm}^{-3}$ – $14.9\times10^9\,\mathrm{cm}^{-3}$. The hydrostatic stratification (e.g., Newkirk 1961) is calculated using a scale height corresponding to a coronal temperature of approximately 1 MK, with a initial density of $n_0 = 10^{10}\,\mathrm{cm}^{-3}$, typical of active–region loops. This treatment naturally produces a gradual density decrease with height, yielding a spatial variation in the local plasma frequency (Figure 2(b)). Therefore, the modeled loop should replicate frequency-drift characteristics similar to SPDBs observations, ensuring consistency between the large-scale plasma environment and observed radio signatures.

Combining the hydrostatic density distribution with the magnetic field strength obtained from the NLFFF model, we calculate the spatial variation of the plasma–to–cyclotron frequency ratio, $\omega_{pe}/\Omega_{ce}$, along the loop (Figure 2(a)). This ratio governs the dominant instability regimes and determines the preferred wave generation and mode conversion characteristics in coronal plasma and loops.

### 3.2. *Selection of Simulation Regions and Electron Injection Parameters*

Based on the derived profile of frequency ratio ($\omega_{pe}/\Omega_{ce}$), three similar representative regions were selected along the magnetic loop, denoted as Boxes A, B, and C (Figure 2(a)). These correspond to frequency ratios of approximately 8, 5, and 2, respectively, tracing the transition from the LT toward the FP. The choice of these locations provides a systematic means to investigate the influence of local plasma conditions on the excitation of wave modes.

The acceleration process responsible for producing the energetic electrons is beyond the scope of the present work. Instead, the energetic electron population injected near the LT region is introduced as an initial condition at the location marked "I" in Figure 2(a), representing electrons that have already been accelerated during flare-associated magnetic reconnection. The initial superthermal electron population is prescribed as a beam-like velocity distribution with a preferred velocity directed along the local magnetic field vector at the injection site. This injection point serves as the starting condition for the GCA calculations, allowing us to analyze electron transport, velocity distribution evolution, and emission properties within the extrapolated magnetic structure. The injected beam distribution conformed to the magnetic field lines (unit vectors of the magnetic field) at the point of injection. A total of two million electrons were released with an average velocity of $0.15c$, corresponding to energies of $\sim 7$ keV. This velocity aligns with the observational interpretation of SPDBs by Zemanová et al. (2020), who point out that the slow positive frequency drifts of SPDBs can possibly be explained by electron beams with velocities $\leq 0.15c$, slower than typical type III electron beams.

The electron timescale of bouncing along the loop is determined by the loop length and the average particle energy. For the present case, with a typical beam speed of $v \approx 0.15c$ and a loop height of approximately $190''$ (corresponding to the extrapolated coronal structure), the average time required for energetic electrons to travel from the LT to the FP is about $\sim 4.5$s. Following impulsive injection, the system is evolved until a sufficient population accumulates at selected observation points along the loop. Accordingly, the VDFs are sampled at three locations separated in the direction of decreasing altitude: Box A at around the LT, Box B at an intermediate position, and Box C near the FP (see Figure 2(a)). The corresponding sampling times are 6s, 8s, and 10s post-injection in boxes A, B, and C, respectively, consistent with the observed $\sim 4$s SPDB duration (see Figure 1). This timing captures the progressive evolution of the electron beam as it propagates downward through regions of increasing background density and stronger magnetic field.

Once the particles injected into a magnetic loop, they will undergo several processes: precipitation based on their pitch angle, reflection due to magnetic mirroring, and scattering due to different types of turbulence. The observed VDFs in different regions reflect the particles' evolution at specific locations and times within the system. The waiting period of few seconds (here is 6s) before capturing VDFs for further PIC simulation is necessary to allow a sufficient number of energetic electrons to accumulate within the observation box. The duration of this waiting period is not predetermined but depends on initial assumptions, such as the mean velocity of the energetic electrons. While a rough estimate suggests $\sim 9$s for traveling the majority of particles from the LT to the FP and return, the actual time is typically shorter because many energetic electrons do not reach the FPs or likely to precipitate if they do. Electrons with small pitch angles propagate faster along magnetic field lines, while those with larger pitch angles are more readily trapped by magnetic mirroring. Pitch-angle scattering was included to mimic the turbulent environment of the loop, with a scattering timescale on the order of seconds, consistent with typical coronal turbulence levels and the timescales of energetic electron evolution observed during solar bursts (Chen & Petrosian 2013). This stochastic process ensures a realistic representation of pitch-angle diffusion and its influence on the evolving VDFs in each region. To optimize the analysis time, VDFs are observed and compared across multiple boxes at different times.

### 3.3. *Particle-in-Cell (PIC) Simulations*

The obtained VDFs from the guiding-center simulations were used as input for fully kinetic PIC simulations conducted with the VPIC code (Bowers et al. 2008, 2009). These simulations allow a detailed examination of wave excitation and mode coupling induced by the nonthermal electron populations derived from the GCA modeling.

In each simulation, the background plasma was modeled as a Maxwellian electron-proton distribution with a temperature of $\sim 1$ MK, corresponding to typical coronal loop conditions. The ratio of energetic to background electron density ($n_e/n_0$) varied along the loop, decreasing from 0.03 at the LT (Box A) to 0.01 near the FP (Box C), reflecting the increasing background density and reduced nonthermal population toward the lower atmosphere. The background magnetic field was aligned with the $z$-axis ($B_0\hat{e}_z$), and the wave vector $\vec{k}$ was set in the $x$–$z$ plane to accommodate both longitudinal and transverse modes. The simulation domain was configured as $1024 \times 1024$ grid cells with a spatial resolution of $\Delta \approx 2.4\lambda_D$, where $\lambda_D$ denotes the electron Debye length. Each cell contained 1000 macroparticles per species to ensure adequate statistical sampling. Charge neutrality is also maintained. The total simulation duration extended to 2000 $\omega_{pe}^{-1}$, allowing sufficient time for the development and saturation of wave instabilities. The temporal and spectral characteristics of the excited modes were analyzed to evaluate the conditions under which fundamental emissions are most efficiently generated.

This integrated modeling framework—combining NLFFF extrapolation, hydrostatic density modeling, guiding-center electron transport, and PIC-based kinetic analysis—provides a physically grounded bridge between SPDB observations and the underlying microphysical processes governing coherent radio emission in solar coronal loops.

## 4. EMISSION RESULTS OF THE EVOLVED DISTRIBUTIONS ALONG THE LOOP

To investigate the wave–particle interaction and subsequent emission characteristics along the coronal loop, we examine the temporal evolution of energetic electron distributions injected at the LT. The injected population possesses a directed beam component with a mean velocity of $\sim 0.15c$, consistent with the characteristic velocities inferred for SPDB-producing electron beams. Due to this initial anisotropy, the electrons preferentially stream along the magnetic field directions, depending on the initial pitch angles and the local magnetic field orientation at the injection site.

The VDFs are constructed by sampling the full particle phase-space within each spatial box. Figure 3 (upper panel) displays the macroscopic guiding-center results, highlighting the evolution of beam-driven structures as electrons move from Box A (close to LT) to Box C (close to FP). At the LT, the distribution exhibits a strong field-aligned beam component, which becomes progressively modified through pitch-angle scattering, magnetic mirroring and precipitation as electrons propagate downward. The stronger magnetic field near the FP restricts the motion of electrons with insufficient parallel energy, resulting in only a subset of electrons satisfying mirror conditions and returning to the LT. This phase-space confinement produces compact VDF structures in $(v_{\parallel}, v_{\perp})$-space, consistent with analytical predictions and earlier numerical studies (e.g., White et al. 1983). The dominant positive gradient in the parallel direction ($\partial f/\partial v_{\parallel} > 0$) establishes favorable conditions for beam-driven instabilities.

To quantify the microphysical instability response, the evolved VDFs are used as input for PIC simulations. The middle panel of Figure 3 compares each VDF to the Maxwellian core of the ambient background plasma after $\sim$ 2000 $\omega_{pe}^{-1}$. In all regions, a clear separation between the energetic tail and the thermal core persists, enabling kinetic free energy to drive wave growth. The corresponding temporal evolution of electromagnetic and particle energy is presented in Figure 3 (bottom panel). The reduction in electron kinetic energy, plotted as $-\Delta E_k(t)$ for clarity, tracks the energy transfer into electromagnetic fields. The $E_z$ component—associated with beam–Langmuir (BL) mode excitation—experiences the earliest and strongest amplification, reaching the maximum energy of $\approx 9 \times 10^{-3} E_{k_0}$, $\approx 2 \times 10^{-3} E_{k_0}$, and $\approx 5 \times 10^{-4} E_{k_0}$ respectively at 600, 700, and 800 $\omega_{pe}^{-1}$ in Boxes A, B, and C, respectively, before saturating.

A Gaussian-filter technique (Yousefzadeh et al. 2021, 2022, 2025) is applied to separate the contributions of individual wave modes. The method involves identifying the $\omega$–$\vec{k}$ regions associated with distinct dispersion branches, reconstructing the energy in physical space via inverse Fourier transform, and integrating spatially to produce mode-specific temporal energy profiles. The resulting wave–vector distribution of $E_z$ is shown in Figure 4(a), revealing prominent BL wave excitation aligned with the beam direction. The strength of this excitation is significantly reduced as we move toward the lower area where the density and strength of magnetic field are increasing. Limited backscattered Langmuir waves are also present but at significantly reduced intensity. The gradual reduction of BL-wave intensity from Box A toward Box C reflects the evolution of the energetic electron population during its transport along the loop. As electrons propagate away from the injection region, precipitation and scattering reduce the number of particles capable of maintaining a positive velocity-space gradient ($\partial f/\partial v_{\parallel} > 0$), thereby decreasing the free energy available for the bump-on-tail instability. Consequently, the growth rate of Langmuir waves becomes weaker toward the lower loop regions. Furthermore, variations in the local plasma and magnetic-field parameters alter the plasma-to-cyclotron frequency ratio, which will further influence the efficiency of wave excitation and subsequent plasma-emission processes.

Figures 4(b)–(c) show the $\omega$–$k$ dispersion analysis including analytical curves for the Langmuir (L) and fundamental (O/F) modes. Across all sampling regions, the dominant energy resides in the BL branch, consistent with beam-driven Langmuir turbulence. Signatures of mode coupling into the fundamental branch appear but remain comparatively weak, and no significant Z -or harmonic- mode activity is observed in these runs. These results suggest that, under SPDB-like parameters, the primary emission mechanism arises from BL waves generated by low-energy beams propagating through a stratified coronal loop, with fundamental emission possible through secondary coupling at later times.

## 5. SUMMARY AND DISCUSSION

In this study, we investigated the generation of plasma emission driven by energetic electrons propagating along a coronal loop and analyzed its connection with SPDBs. While previous numerical work has demonstrated that electron beams can generate plasma emission in weakly magnetized plasmas using analytical beam prescriptions (e.g., Chen et al. 2022), the present study advances this understanding by integrating a realistic three–step numerical modeling pipeline—NLFFF extrapolation, guiding–center electron transport, and PIC simulations—with an observational SPDB event. However, electron beams injected into coronal structures are inherently unstable and undergo substantial evolution, including the development of turbulence and deformation of their VDFs. This naturally raises the question of whether these evolved VDFs retain the ability to excite plasma waves and generate radio emission. This combined approach enables us to connect the kinetic–scale emission physics to macroscale coronal loop structure derived from solar data, making this a preliminary effort in bridging the gap between idealized beam–plasma theory and flare–associated radio burst observations.

The injected electron beams at the LT evolve through continuous interaction with the ambient coronal plasma, undergoing magnetic mirroring, pitch–angle scattering, turbulence development, and partial precipitation into the lower atmosphere. The characteristic timescale for this evolution can be estimated from the loop geometry and the mean energy of the accelerated electrons. This timescale defines the interval over which the majority of particles complete at least one bounce cycle, allowing them to undergo significant interaction with the loop environment before emission properties are evaluated. Therefore, we have considered the evolved velocity distribution at the LT location (Box A) at $t \sim 6\,\mathrm{s}$, followed by the distribution at an intermediate height (Box B) at $t \sim 8\,\mathrm{s}$, and near the FP (Box C) at $t \sim 10\,\mathrm{s}$. This $4\,\mathrm{s}$ temporal separation between analysis locations corresponds to the characteristic duration of the observed SPDB signatures in the dynamic spectrum and reflects the approximate time delay associated with frequency–dependent emission along the loop. This selection ensures that our modeled VDFs reflect physically meaningful, post–evolution states that are temporally consistent with the observed SPDB evolution.

The selection of a representative post-evolution state is supported by the observation that the VDF morphology remains quasi-stable over characteristic times of $\sim 0.3\,\mathrm{s}$ for beams with initial velocity $\sim 0.24c$ (Yousefzadeh et al. 2025), and slightly longer for the lower initial beam velocity ($\sim 0.15c$) considered here. Thus, the chosen interval captures a physically meaningful phase in which the beam has partially relaxed yet remains capable of driving wave excitation. Furthermore, our simulations show that plasma emission can occur not only at a selected time (e.g., $6\,\mathrm{s}$ at the LT) but also within a brief temporal window ($\sim 0.5\,\mathrm{s}$). This is consistent with the characteristic persistence of EVDF structures and the observed SPDB durations across different frequencies, because the VDF remains relatively stable during this short period.

Our results reveal that the evolved beams predominantly excite Langmuir waves and fundamental plasma emission at all three analyzed locations along the loop. Due to the relatively small injection energy ($\sim 0.15c$), the beam strength is insufficient to generate significant harmonic radiation, in contrast to previous work employing higher-energy electrons. Although emission occurs throughout the loop—from the loop-top toward the footpoint—the intensity decreases progressively toward lower altitudes. This trend likely reflects the combined effects of increasing magnetic field strength, decreasing the relative density of energetic electrons per background one, and changing plasma parameters (e.g., reduced frequency ratio) near the loop footpoints, all of which act to damp beam-driven instabilities.

The present study is intentionally constrained by observational properties of the selected SPDB event. The analyzed magnetic loop was chosen based on the temporal correspondence between the radio emission and UV/EUV flare kernels, while the loop geometry, source altitude, density structure, and electron-beam velocity were selected to remain consistent with observational estimates. The three analyzed locations correspond to plasma frequencies spanning the observed SPDB frequency range of approximately 600–1100 MHz, while the temporal sequence of the selected EVDFs follows the observed slow frequency-drift evolution.

It should be emphasized that the purpose of the simulations is not to reproduce these observational constraints independently, but rather to determine the physical emission processes that arise under conditions inferred from the observations. The PIC simulations demonstrate that the evolved electron distributions retain sufficient free energy ($\partial f/\partial v_{\parallel} > 0$) to drive beam-plasma instabilities, resulting in the excitation of BL waves and subsequent fundamental plasma emission. The dispersion analysis further confirms the presence of wave modes consistent with the plasma-emission mechanism. The modeled emission strength decreases from the loop-top toward the footpoint, reflecting the gradual evolution of the energetic electron population along the loop. Although the current implementation provides only three discrete emission points rather than a continuous synthetic dynamic spectrum, the modeled temporal evolution reproduces the observed frequency-drift sequence and demonstrates that evolved electron distributions can efficiently generate BL waves and fundamental plasma emission throughout the loop. Future extensions of this approach, employing a larger number of sampling locations and PIC simulations, may enable the construction of fully synthetic dynamic spectra for direct comparison with observations.

Overall, this study establishes a coherent framework linking coronal loop topology, hydrostatically stratified plasma properties, evolved electron distributions, and plasma wave excitation to the generation of observed SPDBs. While harmonic emission remains weak under the beam energies explored here, the demonstrated production of fundamental plasma radiation and its agreement with temporal SPDB signatures suggests that plasma emission is a likely mechanism for these bursts. Future work incorporating higher beam energies, other types of radio bursts, multi-scale turbulence, and full mode-conversion processes will further clarify the relative roles of fundamental and harmonic emission channels in fine-structured solar radio bursts. In parallel, near-Sun observations from Solar Orbiter (SoLO) and future missions such as Solar Close Observations and Proximity Experiments (SCOPE) may provide additional constraints on the near-Sun plasma environment, energetic-electron transport, and wave activity relevant to beam-driven radio emission close to the Sun (Lin et al. 2025). Although such missions do not directly resolve SPDB source regions, they can help test the key physical ingredients assumed in multiscale models such as the one developed here, thereby linking the proposed three-step framework to future observational validation.

ACKNOWLEDGMENTS

The research results are sponsored and supported mainly by the National Key R&D Program of China No. 2022YFF0503804. M.Y. acknowledge supports from the Youth Specialty Category grant under the project No. ZR2024QD206. M.K., A.K. and A.Z. acknowledge institutional support from project RVO-67985815. M.K. and A.K. also acknowledge support from the project GAČR grant 26-20555L. The authors acknowledge the open-source VPIC code by Los Alamos National Labs (LANL), the Beijing Super Cloud Computing Center (BSCC)[7] for providing computational resources and the use of the CALLISTO data by the Institute for Data Science FHNW Brugg/Windisch (Switzerland)[8]. M.Y. gratefully acknowledges Prof. Chengcai Shen (YNAO) for his helpful comments and discussion.

[7] http://www.blsc.cn/

[8] https://www.e-callisto.org/Data/data.html

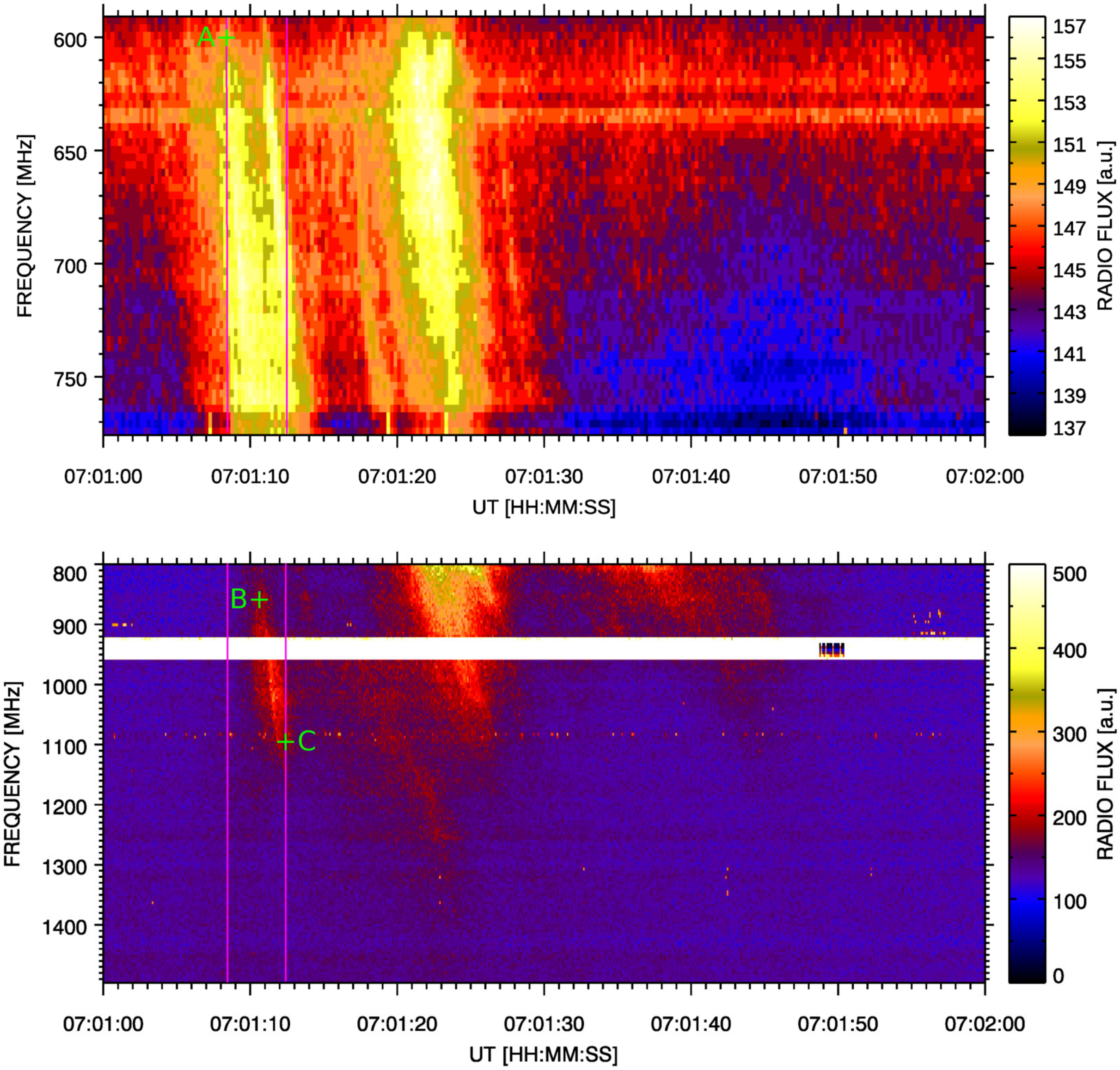

**Figure 1.** Dynamic spectra from the Ondřejov and BLEN7M Callisto radiospectrographs. The BLEN7M Callisto and RT5 radio spectra (upper and lower panels) show bursts from the SPDB group only in the time range 07:01–07:02 UT. The 4 s duration of the selected SPDB is indicated by purple vertical lines in the spectra. Points A, B, and C in the spectra (marked by green crosses) correspond to locations along the coronal loop where the radio emission was calculated. The radio fluxed are presented in relative units.

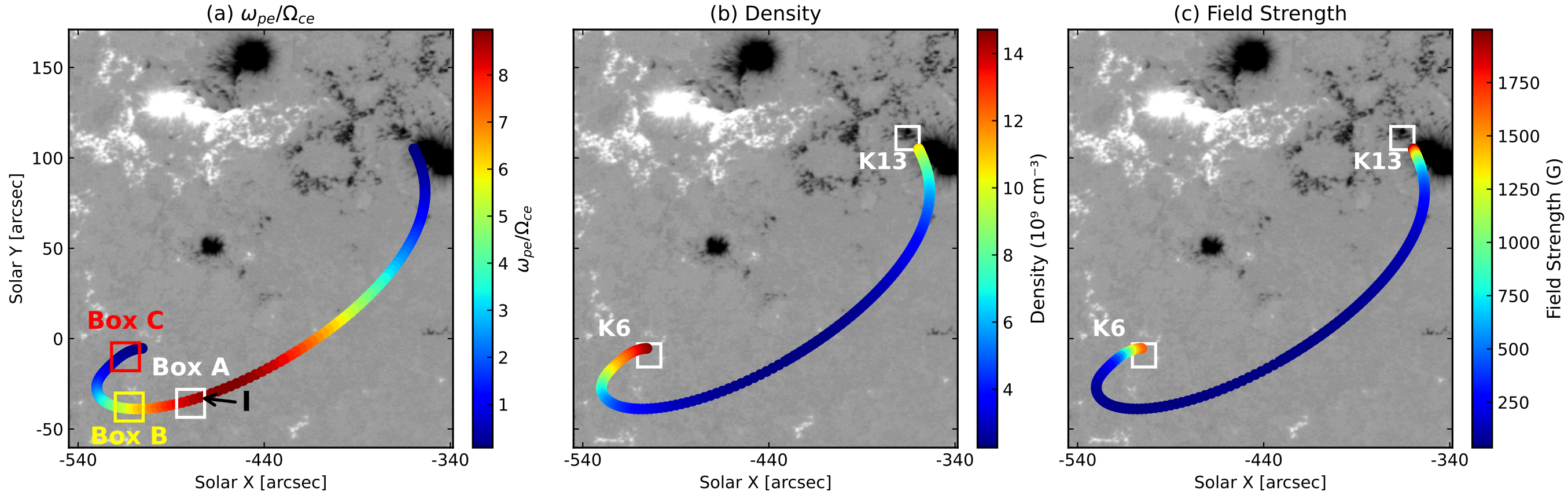

**Figure 2.** (a) Reconstructed magnetic field line from the NLFFF extrapolation, used for deriving the frequency ratio profile along the selected loop. (b) Hydrostatic density model along the magnetic field line consistent with the observed radio-emission frequency range and SPDB dynamic spectrum. (c) The extrapolated field line of the coronal loop indicates the probable emission region, with color representing magnetic field strength in Gauss. The letter "I" marks the injection site. The approximate locations of K13 and K6 flare kernels are shown on the background HMI image ((b) and (c)). The magnetogram data was observed at 07:00 UT.

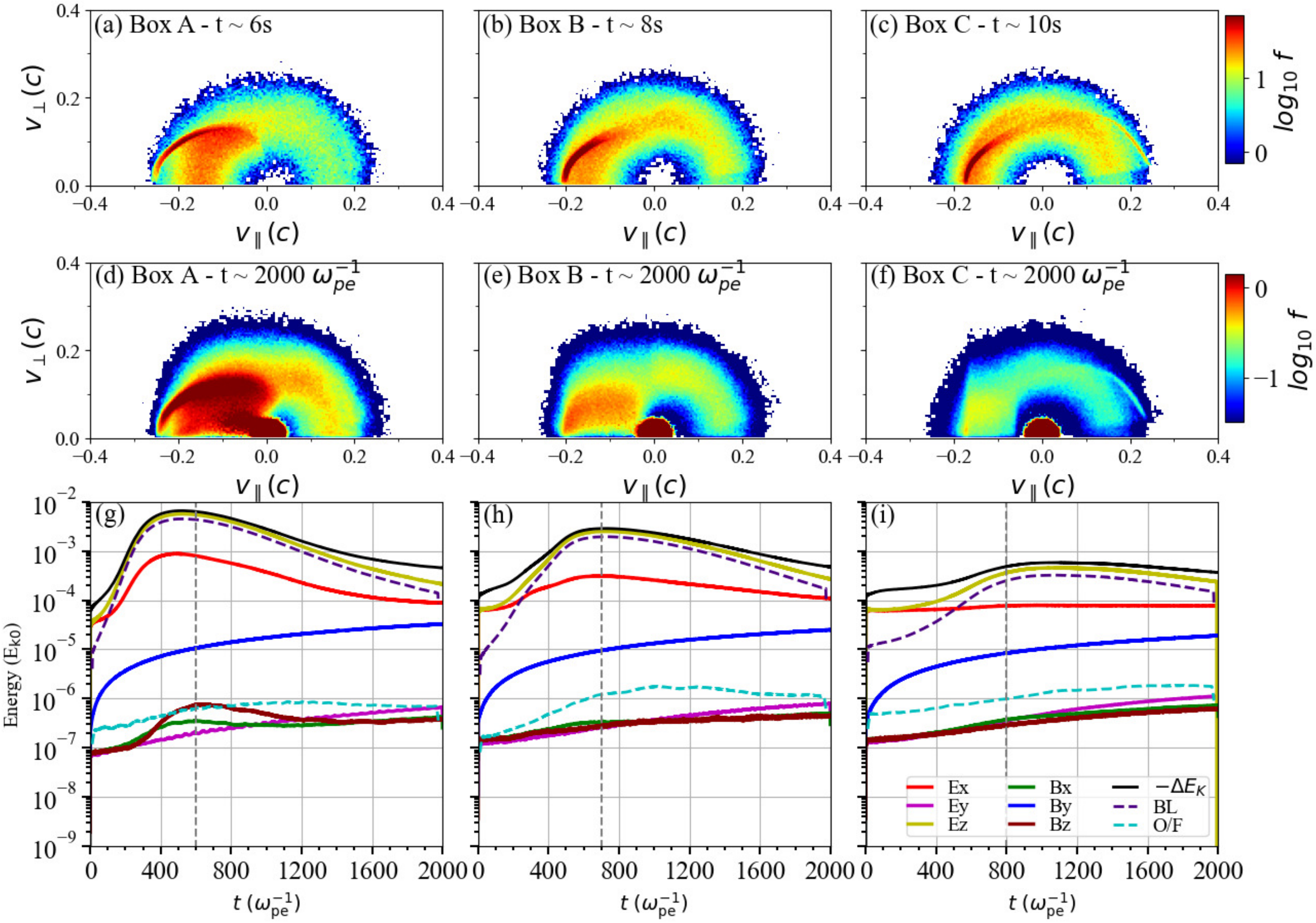

**Figure 3.** Macroscopic guiding-center results show the evolution of beam-driven structures as electrons move from Box A (t∼ $6s$) to Box B (t∼ $8s$) and Box C (t∼ $10s$), representing the evolution of energetic electrons from the looptop to a footpoint (a-c). The interaction of the energetic electron VDFs with the background plasma after around $\sim 2000\ \omega_{pe}^{-1}$ (d-f). Temporal evolution of electromagnetic and particle energy variation for the six field components and the negative change in total electron kinetic energy ($-\Delta E_k(t)$), which shows the energy mostly transferred to the parallel electric fields (g-i). The vertical lines in different cases indicate the time where the parallel electric field reached its maximum energy. The animation shows the evolution of the EVDFs in different boxes, starting at injection time and continuing for approximately 12 seconds. The real-time duration of the video is 24 s.
(An animation of this figure is available in the online article.)

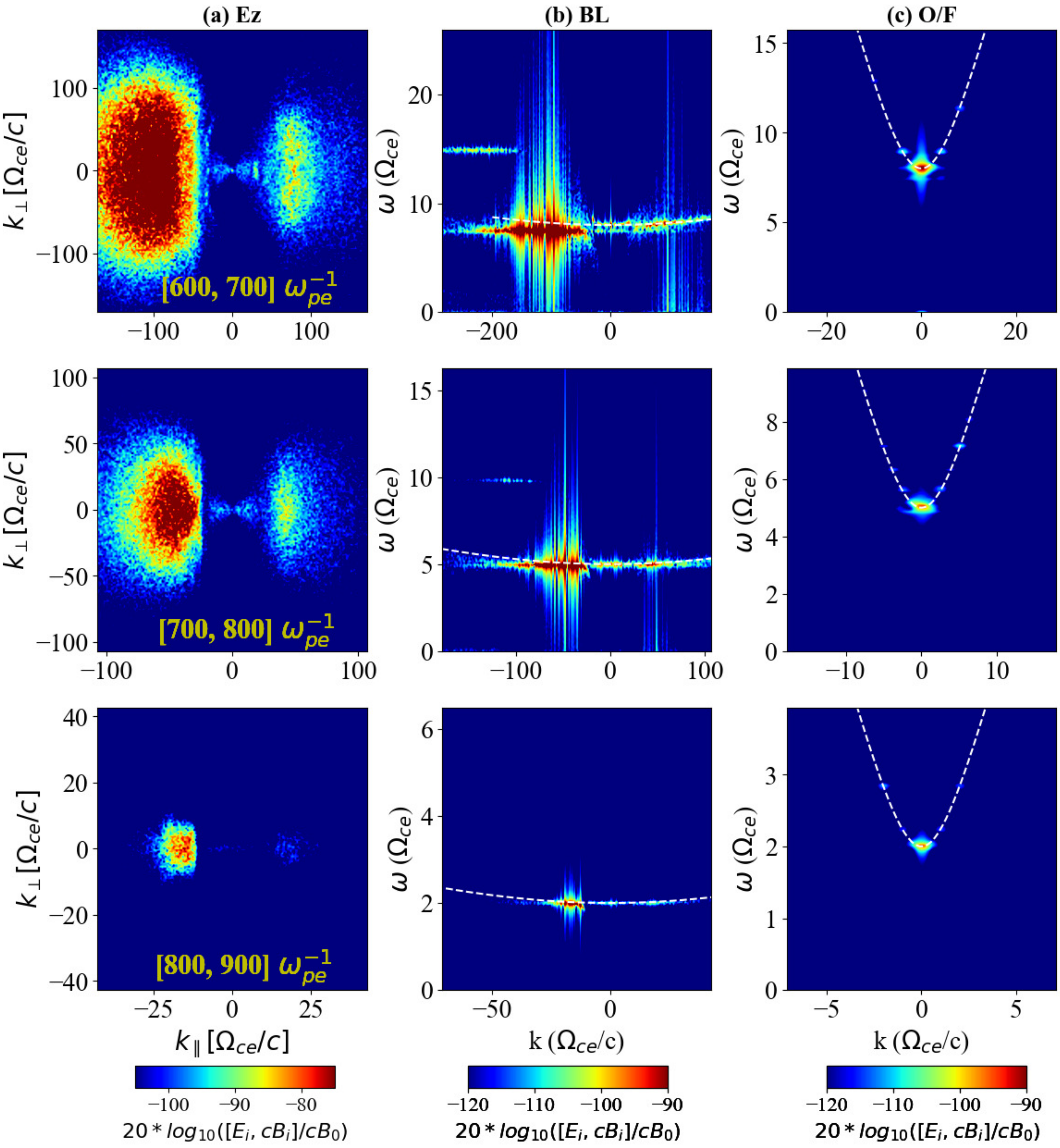

**Figure 4.** (a) Wave-vector distribution of $E_z$, showing prominent beam-driven Langmuir (BL) wave excitation and limited backscattered Langmuir waves in $k_\parallel, k_\perp$ space. (b)-(c) $\omega$–$k$ dispersion analysis with analytical curves for Langmuir (L) and fundamental modes (O/F), indicating dominant energy in the BL branch and signatures of mode coupling into the fundamental. The dashed lines illustrate dispersion curves, which were derived from magnetoionic theory. Rows represent results for Box A (top), Box B (middle), and Box C (bottom).